\documentclass{aa}

\usepackage{graphics}
\begin{document}

\thesaurus{06(08.22.3; 08.12.1; 08.16.4; 08.01.3)}

\title{A comparison of light and velocity variations in Semiregular variables}

\author{T.~Lebzelter\inst{1}
\and L.L.~Kiss\inst{2}
\and K.H.~Hinkle\inst{3}}

\offprints{lebzelter@astro.univie.ac.at}

\institute{Institut f\"ur Astronomie, Universit\"at Wien, T\"urkenschanzstr.\,17, A-1180 Vienna,
Austria
\and Department of Experimental Physics and Astronomical Observatory, University of Szeged, Hungary
\and
Kitt Peak National Observatory, National Optical Astronomy 
Observatory
\thanks{Operated by the Association of Universities for Research
in Astronomy, Inc.~under cooperative agreement with the National Science
Foundation},
950 N.Cherry Avenue, P.O.Box 26732, Tucson, Arizona 85726}

\date{Received / Accepted}

\titlerunning{Light and velocity variations in SRVs}

\maketitle

\begin{abstract}
NIR velocity variations are compared with simultaneous
visual light curves for a sample of late-type semiregular variables (SRV). 
Precise radial velocity measurements are also presented for the SRV V450
Aql. Our aim is to investigate the nature of the irregular light
changes found in these variables.

Light and velocity variations are correlated in all stars of
our sample. Based on these results we discuss several possibilities
to explain the observed behavior. We find that pulsation is responsible
for large amplitude variations. In a recent paper Lebzelter (\cite{Lebzelter99})
invoked
large convective cells to understand observed velocity variations. This possibility
is discussed with respect to the observed correlation between light
and velocity changes. In the light of these results we investigate
the origin of the semiregular variations.
\keywords{stars: variables: general -- stars: late-type -- stars: AGB and post-AGB --
stars: atmospheres} \end{abstract}

\section{Introduction}

The first discovery of semiregular light variation for a red giant
($\mu$ Cep) dates to 1782 (see Burnham \cite{Burnham78} for a historical
review). Since that time an enormous
amount of light curve data has been collected, with most of 
the data for large amplitude late-type variable (miras) 
coming from amateur astronomers.  For the lower amplitude late-type
variables a distinct group has been assigned, the semiregular
variables (SRVs; Kholopov et al.~\cite{GCVS4}, GCVS4). 

In this paper we
will focus on the semiregular subclasses SRa and SRb only, and exclude
SRc (supergiants) and SRd (yellow giants).  A comprehensive discussion
of the properties of SRas and SRbs has been given by Kersch\-baum \& Hron
(\cite{KH92}, \cite{KH94}). The SRVs are evolved stars, probably
located on the AGB; however their evolutionary status on the AGB is
still a matter of debate (e.g.~Lebzelter \& Hron \cite{LH99}).  The
bluer SRVs, which typically also have the shorter
periods, show less mass loss and no indications for thermal pulses. The
redder SRVs have longer periods, and can show mass
loss comparable to the miras.

The cause of the irregularities in the light variations of the SRVs is far from
being understood.  Recently, empirical evidence has
been presented for multiple period changes (Mattei et
al.~\cite{MFH98}, Kiss et al.~\cite{KSCM99a}), which may be due to many
(2 or more) simultaneously excited modes of pulsation. The observed
mode changes in several semiregular stars (Cadmus et al.~\cite{CWSM91},
Percy \& Desjardnis \cite{PD96}, Bedding et al.~\cite{BZJF98}, Kiss et
al.~\cite{KSCM99a}) strongly suggest a pulsational
origin for the multiperiodic nature. However, other explanations,
e.g. chaotic phenomena (Aikawa \cite{Aikawa87}, Icke et al.~\cite{IFH92}) 
perhaps in combination with an interplay of excited
modes of pulsation (Buchler \& Goupil \cite{BG88}) or coupling of
rotation and pulsation (Barnbaum et al.~\cite{BMK95}) have been
suggeested in the literature. One has to be
aware that all of these mechanisms may simply be cases
of `chaos' (see e.g.~Buchler \& Koll\'ath \cite{BK00}). Further explanations
include
dust-shell dynamics (H\"ofner et al.~\cite{HFD95}) or
large convective cells (Antia et al.~\cite{ACN84}).
Observed spatial asymmetries in
some Mira variables (Karovska \cite{Karovska99}) could imply the
possibility of non-radial oscillations.  This was also discussed as an
explanation for the amplitude modulation of the SRb-type
variable RY~UMa (Kiss et al.~\cite{KSSM99b}).

A problem in understanding the nature of the SRVs is that nearly 
all observations have been of visual colors.  To develop a physical
model for the SRV changes, monitoring of
time variability of other parameters like velocity or temperature
variations is needed.  One approach is to monitor the 
velocity variations of SRVs 
closer to the pulsation driving zone through time-series high-resolution 
infrared spectroscopy.
Hinkle et al.~(\cite{HLS97}) and Lebzelter (\cite{Lebzelter99}) reported
on time-series high-resolution spectroscopy of 
second overtone and high
excitation first overtone CO lines located
in an opacity minimum around 1.6\,$\mu$m.  
However, in both papers only
phase information (relative to a close light curve maximum) was used 
to search for regularities in the
velocity variations and to derive an amplitude of the velocity curve.
As the variations in SRVs are far from being regular both in the sense
of period length as well as amplitude, a phase estimated in this way
may significantly misrepresent the behavior of the light curve.

A direct comparison between light and velocity changes has been done
for the mira variable \object{$\chi$ Cyg} (Hinkle et al.~\cite{HHR82}). 
The velocity curves of miras are all similar and
discontinuous with an amplitude of 20 to 30 km/s. Line doubling is
observed around maximum light, and at that phase typically both maximum
and minimum velocity occur. Around phase 0.4, i.e.~shortly before light
minimum, the velocity curve reaches the center-of-mass velocity. This
point in the variability cycle might therefore be related to the
maximum expansion of the layer producing the CO lines used to derive
the velocity curve. Other miras show similar results; 
the velocity amplitudes of SRVs are significantly less 
(Hinkle et al.~\cite{HSH84}, Hinkle et al.~\cite{HLS97}). 

This paper is dedicated to the {\it irregularities}.   We
will compare simultaneously measured light and 
velocity curves to correlate the irregularities. The velocity 
variations represent the
movement of atmospheric layers due to stellar pulsation. In this way we
will investigate whether the irregularities in the light curve are
due to nonregular pulsation or due to surface structures like active
regions or velocity fields introduced by convective cells.

\section{Observations} 

\subsection{Velocity variations} 
Most of the velocity data of SRVs that will be discussed 
have been published elsewhere (Hinkle et al.~\cite{HLS97}, Lebzelter
\cite{Lebzelter99}). We selected stars that have a reasonable
number of velocity and light curve measurements.  
For comparison we have also included the short period mira RT Cyg. This star
has a period similar to the SRVs but with a significantly larger
amplitude in V. The velocity amplitude is also similar to that of other
miras (Lebzelter et al.~\cite{LHH99}).
The final sample covers a large range in period and amplitude. Both
types, SRa and SRb, are included.
Table\,\ref{tab1}
gives an overview on the origin of the data and the instruments used to
obtain them as well as fundamental properties of the stars.

\begin{flushleft} 
\begin{table*}
\caption[]{Properties of stars selected for the present investigation.
Column 3, 4, 5 and 6 give general properties of the objects from the
GCVS4. Column 7 and 8 list the source of the velocity data, where CF
stands for the Coud\'e Feed telescope on Kitt Peak. References are 
encoded as follows: HLS97 = Hinkle et al.~1997; LHH99 = Lebzelter
et al.~1999; LZ99 = Lebzelter 1999. \label{tab1}}
\begin{tabular}{l|l|rrrr|cc} 
\hline 
Object & IRAS &\multicolumn{4}{c}{GCVS data} & \multicolumn{2}{c}{velocity data}\\
 & name & type & spectrum & period [d] & V ampl.~[mag] & reference & instrument\\ 
\hline 
\object{Aql V450} & 19312+0521 & SRb & M5 & 64.2 & 0.35 & this paper & FTS KPNO\\ 
\object{Cas SV} & 23365+5159 & SRa & M6 & 265 & 3.4 & HLS97 & FTS KPNO\\ 
\object{Cyg W} & 21341+4508 & SRb & M4-6 & 131 & 2.1 & HLS97 & FTS KPNO\\ 
\object{Cyg RT} & 19422+4839 & M & M2-8.8 & 190 & 7.1 & LHH99 & FTS KPNO\\ 
\object{Cyg RU} & 21389+5405 & SRa & M6e & 233 & 1.4 & HLS97 & FTS KPNO \\ 
\object{Dra TX} & 16342+6034 & SRb & M4e-5e & 78 & 2.3 & LZ99 & CF + NICMASS \\ 
\object{Her g} & 16269+4159 & SRb & M6 & 89 & 2.0 & LZ99 & CF + NICMASS\\ 
\object{Her ST} & 15492+4837 & SRb & M6-7 & 148 & 1.5 & LZ99 & CF + NICMASS\\ 
\object{Hya W} & 13462-2807 & SRa-M & M8e & 382 & 3.9 & HLS97 & FTS KPNO\\ 
\hline \end{tabular} \end{table*}
\end{flushleft}

FTS spectra were obtained over a broad wavelength interval always
covering the entire 1.6\,$\mu$m H window and frequently the 
entire 1.6-2.5\,$\mu$m H and K region.  We present a single velocity
from these measurements.
The Coud\'e Feed spectrograph with the NICMASS detector covers a very 
small spectral region,
about 2 percent of the H band window (Lebzelter \cite{Lebzelter99}).
The Coud\'e Feed observations were always obtained
in two wavelength regions, both around 1.6\,$\mu$m, that were analyzed
separately.  In the figures of this paper these two sets of NICMASS
observations will be marked by two
different symbols. 

Previously unpublished velocity data for the
semiregular variable V450 Aql are also presented. Observations were
obtained with the
FTS at the 4\,m Mayall telescope on Kitt Peak. The data are listed in
Table\,\ref{tab2}. Details on the data reduction techniques can be
found in Hinkle et al.~(\cite{HHR82}).  The excitation temperature and
column density were derived from the CO
lines for two observations as described in Hinkle et al.~(\cite{HHR82}). 
As has been found for other similar SRVs, the uncertainty of these
numbers (T$_\mathrm{exc}$=3300$\pm$200 and log(NL[CO])=22.8$\pm$0.1) is too large 
to allow derivation of stellar variations in low amplitude variables.

Some confusion is found in the literature concerning the spectral type
of V450 Aql.  Most authors agree that the spectral type is M5 to
M5.5III (e.g.~Keenan \& McNeil \cite{KM89}), while the Hipparcos Input
Catalogue lists M8V, apparently based on a measurement by McCuskey
(\cite{McCuskey49}). From its IRAS colours V450 Aql clearly belongs to
the blue SRVs (Kerschbaum \& Hron \cite{KH92}). 

\begin{flushleft} 
\begin{table} 
\caption[]{FTS measurements for V450
Aql. Spectra have been apodized by Norton \& Beer (\cite{NB76})
function I2. Resolution (theor.~FWHM) of the spectra is 0.07\,cm$^{-1}$. \label{tab2}}
\begin{tabular}{lcccc} 
\hline 
JD & S/N & integration & radial & number\\ 
2440000+ & & time [min] & velocity [km/s] & of lines\\
\hline 
5604.7 & 76 & 110 & $-$52.34$\pm$0.14 & 76\\ 
5752.3 & 61 & 34 & $-$52.98$\pm$0.12 & 74\\ 
5776.2 & 48 & 42 & $-$52.90$\pm$0.10 & 75\\ 
5782.2 & 62 & 34 & $-$52.72$\pm$0.11 & 74\\ 
5799.1 & 69 & 51 & $-$53.79$\pm$0.12 & 74\\ 
5812.1 & 52 & 34 & $-$53.62$\pm$0.15 & 76\\ 
5837.1 & 58 & 56 & $-$53.63$\pm$0.11 & 78\\ 
5841.0 & 43 & 50 & $-$53.89$\pm$0.09 & 79\\ 
6005.5 & 52 & 50 & $-$52.32$\pm$0.16 & 76\\ 
6018.4 & 85 & 67 & $-$52.51$\pm$0.12 & 73\\ 
\hline
\end{tabular} \end{table} \end{flushleft}

The literature values
for the radial velocity of V450 Aql scatter between $-$52\,km/s
(Feast, Woolley \& Yilmaz \cite{FWY72}) and $-$50.6\,km/s (Jones \&
Fisher \cite{JF84}). We will use an average value of $-$51.3\,km/s. All
velocity measurements we obtained for this object (Table\,\ref{tab2})
are slightly below this literature value.  V450 Aql therefore exhibits
the same behavior found already for a number of SRVs by Lebzelter
(\cite{Lebzelter99}), namely a systematic shift between the literature values
and the measured CO velocities.

\subsection{Light curve data} 
The measurement and long-term monitoring of long-period variable magnitudes 
is one field in astronomy where
the contribution of amateur astronomers is crucial. The investigation
presented in this paper was made possible by the rich data base of
individual magnitude estimates provided by associations of amateur
astronomers.  Three sources of visual
data were utilized, namely those of the Association Francaise des Observateurs
d'Etoiles Variables (AFOEV\footnote{\tt
ftp://cdsarc.u-strasbg.fr/pub/afoev}), the Variable Star Observers'
League in Japan (VSOLJ\footnote{\tt
http://www.kusastro.kyoto-u.ac.jp/vsnet/gcvs}) and the American
Association of Variable Star Observers (AAVSO International Database
which includes part of AFOEV and VSOLJ observations). The possible
effects of overlapping data series (i.e.~the common estimates can be 
included twice or thrice) were avoided, because where the (most complete) AAVSO 
data were available we did not mix them with the other ones. On the
other hand, there are practically no common observations in the AFOEV and
VSOLJ data base.

In order to decrease the effects of the observational scatter, we
applied similar data handling as used in Kiss et al.~(\cite{KSCM99a}). That
means data averaging in 5- or 10-day bins depending on the actual
period value. Further noise-filtering was done by applying a Gaussian
smoothing with a FWHM of 4 and 8 days (80 percent of the bin). This
was necessary in the very low-amplitude regime, where the usual
accuracy of individual estimates (about $\pm$0.3 mag) would obscure the
light variations. The use of this smoothing was illustrated in a
comparison with simultaneous photoelectric measurements (see Fig.\,2 in
Kiss et al.~\cite{KSCM99a}) and turned out to be very effective in the
case of light curves with numerous and densely distributed data. The
estimated precision (in the {\it visual} system) is somewhat better
than 0.1 mag. However, since there are no independent and
simultaneous well-calibrated photometric data, the accuracy can only be estimated.

\section{Results}
\begin{figure} \resizebox{\hsize}!{\includegraphics{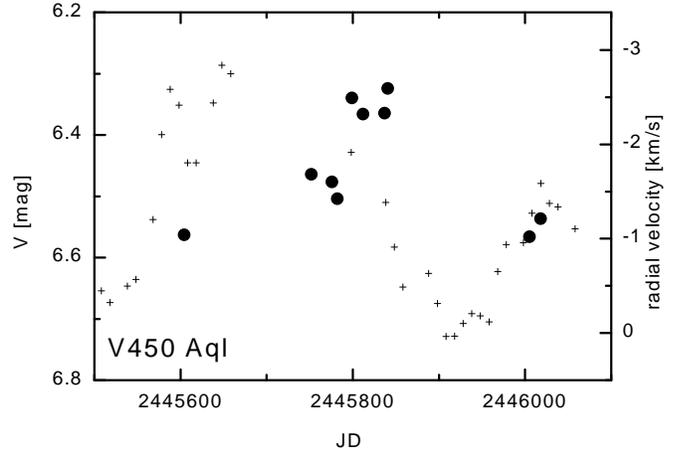}}
\caption[]{Light and velocity variations of V450 Aql. Filled circles
mark velocity measurements (right axis). No error bars are given
as the error bar would be typically of the size of the symbols
as can be easily seen from Table~\ref{tab2}. The stellar velocity ($-$51.3\,km/s
from blue/optical spectra, see text)
has been
subtracted. To allow a better comparison velocity axis is in reverse
order. Light curve data points are labeled with crosses. \label{fig1}}
\end{figure}

\begin{figure} \resizebox{\hsize}!{\includegraphics{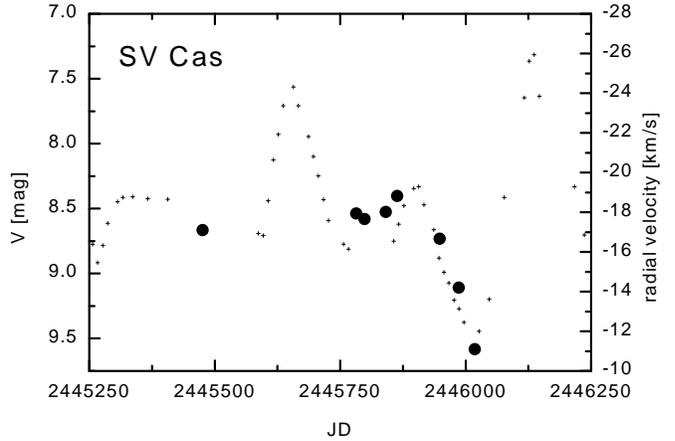}}
\caption[]{Light and velocity variations of SV Cas. For this star no
center-of-mass velocity is available in the literature. Therefore
absolute velocity values are plotted.  Same symbols as in Fig.~
\ref{fig1}. \label{fig2}} \end{figure}

\begin{figure} \resizebox{\hsize}!{\includegraphics{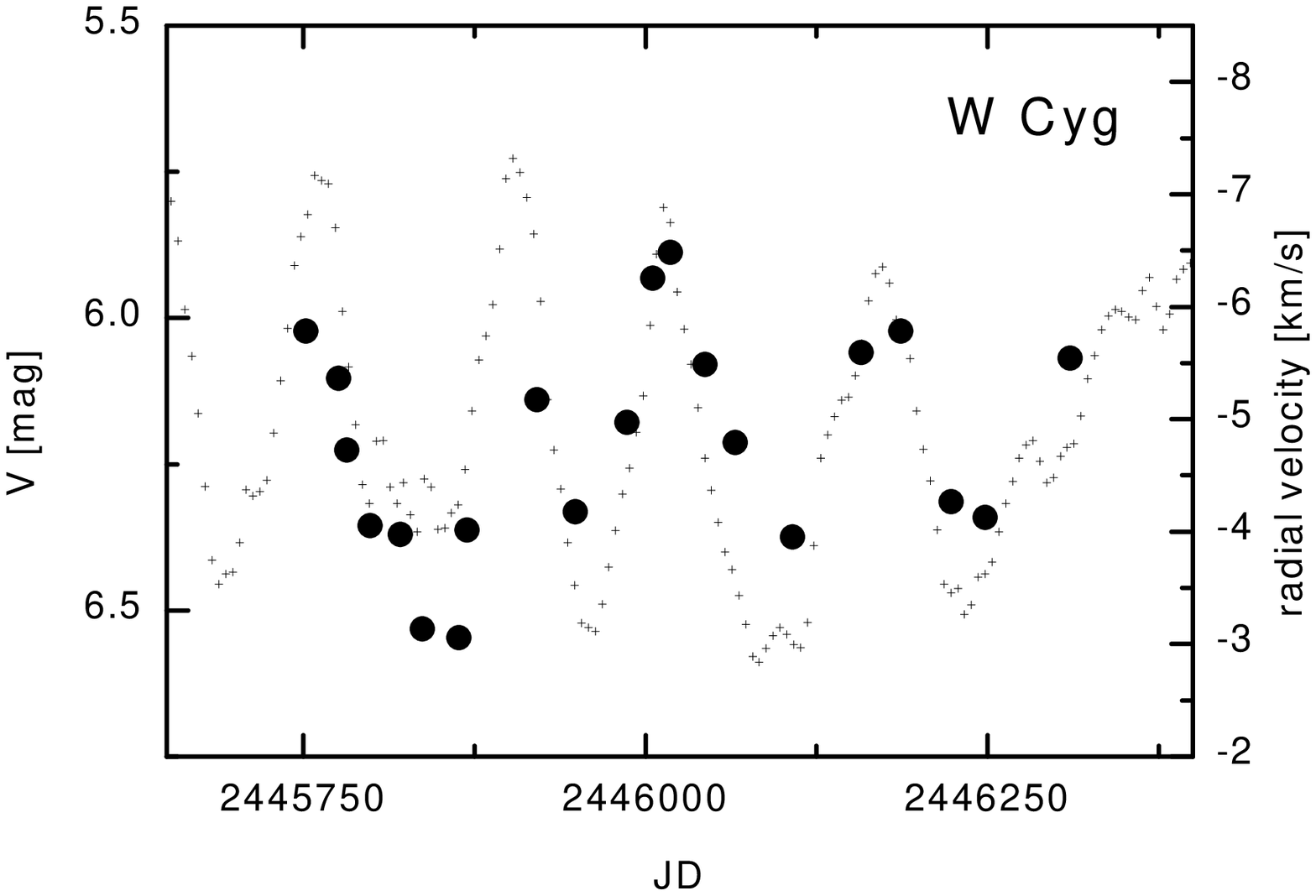}}
\caption[]{Light and velocity variations of W Cyg. Stellar velocity 
($-$14.4\,km/s from thermal microwave CO, see Hinkle et al.~\cite{HLS97}) has
been subtracted from the velocity data. Same symbols as in Fig.~
\ref{fig1}. \label{fig3}} \end{figure}

\begin{figure} \resizebox{\hsize}!{\includegraphics{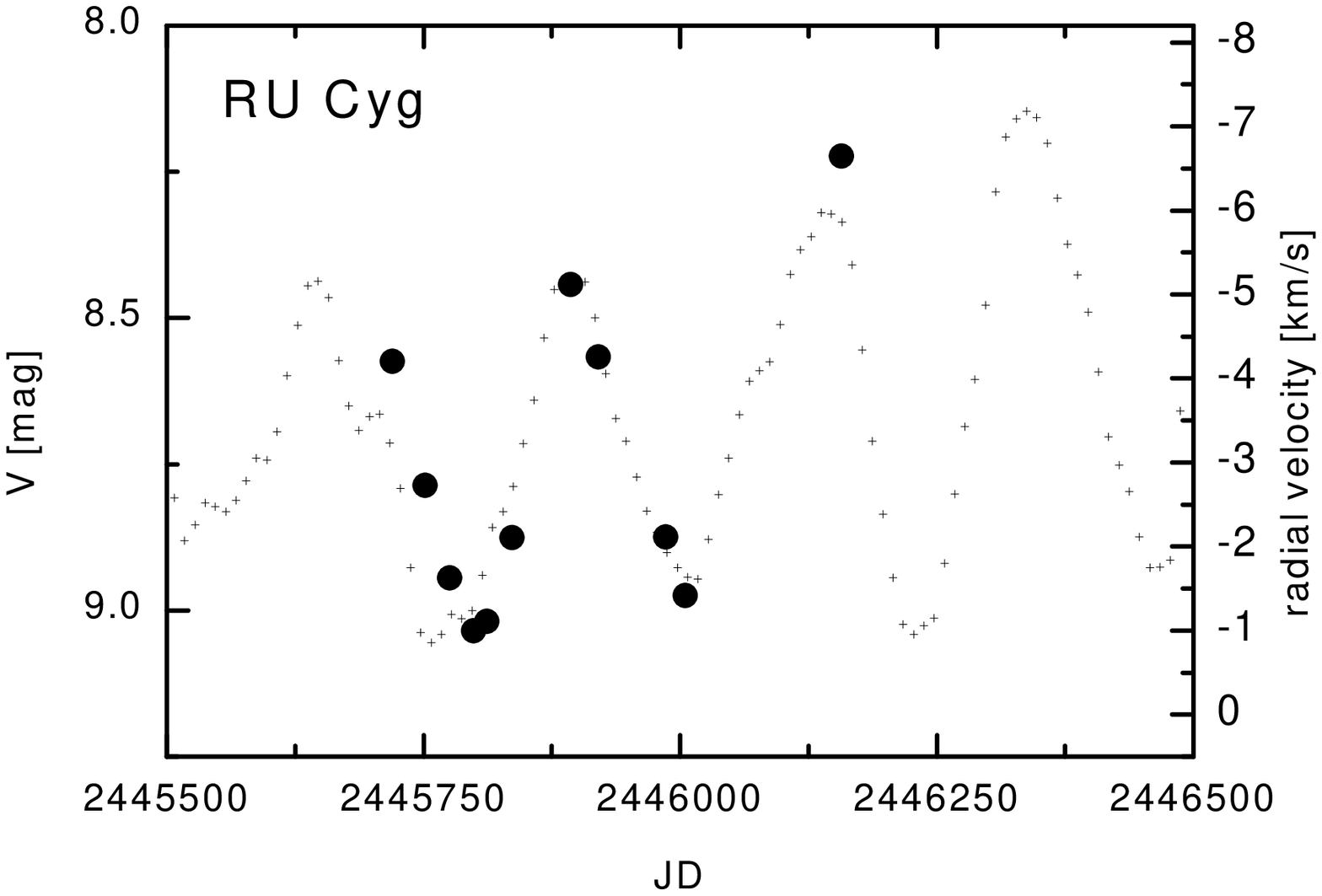}}
\caption[]{Light and velocity variations of RU Cyg. Stellar velocity
($-$6.3\,km/s from thermal microwave CO, see Hinkle et al.~\cite{HLS97})
has been subtracted from the velocity data. Same symbols as in Fig.~
\ref{fig1}. \label{fig4}} \end{figure}

\begin{figure} \resizebox{\hsize}!{\includegraphics{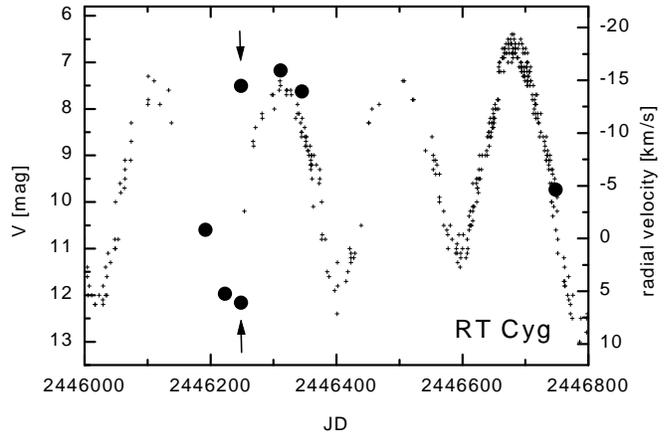}}
\caption[]{Light and velocity variations of the short period mira RT
Cyg.  Stellar velocity 
($-$118\,km/s from blue/optical spectra, see Lebzelter et al.~\cite{LHH99})
has been subtracted from the velocity data. The
arrows indicate the phase of line doubling.  Same symbols as in Fig.~
\ref{fig1}. \label{fig5}} \end{figure}

\begin{figure} \resizebox{\hsize}!{\includegraphics{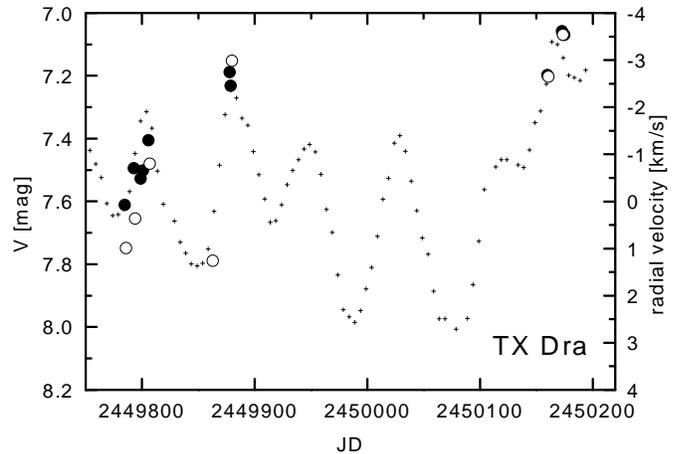}}
\caption[]{Light and velocity variations of TX Dra. Stellar velocity
(46.4\,km/s from blue/optical spectra, see Lebzelter \cite{Lebzelter99})
has been subtracted from the velocity data. Velocity has been measured
in two wavelength regions (see text). Wavelength region\,1 is marked by
filled circles, wavelength region\,2 by open circles. \label{fig6}}
\end{figure}

\begin{figure} \resizebox{\hsize}!{\includegraphics{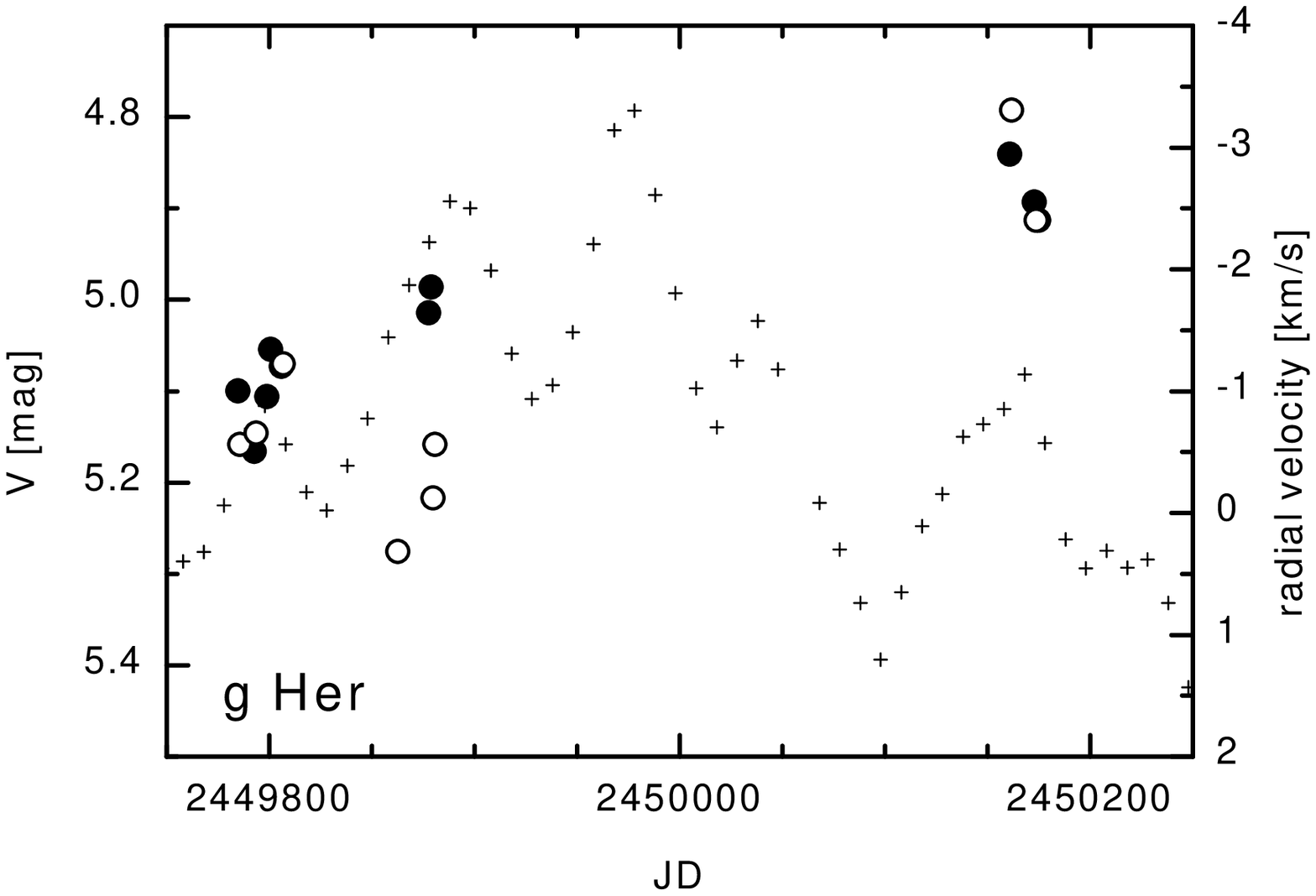}}
\caption[]{Light and velocity variations of g Her. Stellar velocity 
(1.4\,km/s from thermal microwave CO, see Lebzelter \cite{Lebzelter99}) has
been subtracted from the velocity data. Same symbols as in Fig.~
\ref{fig6}. \label{fig7}} \end{figure}

\begin{figure} \resizebox{\hsize}!{\includegraphics{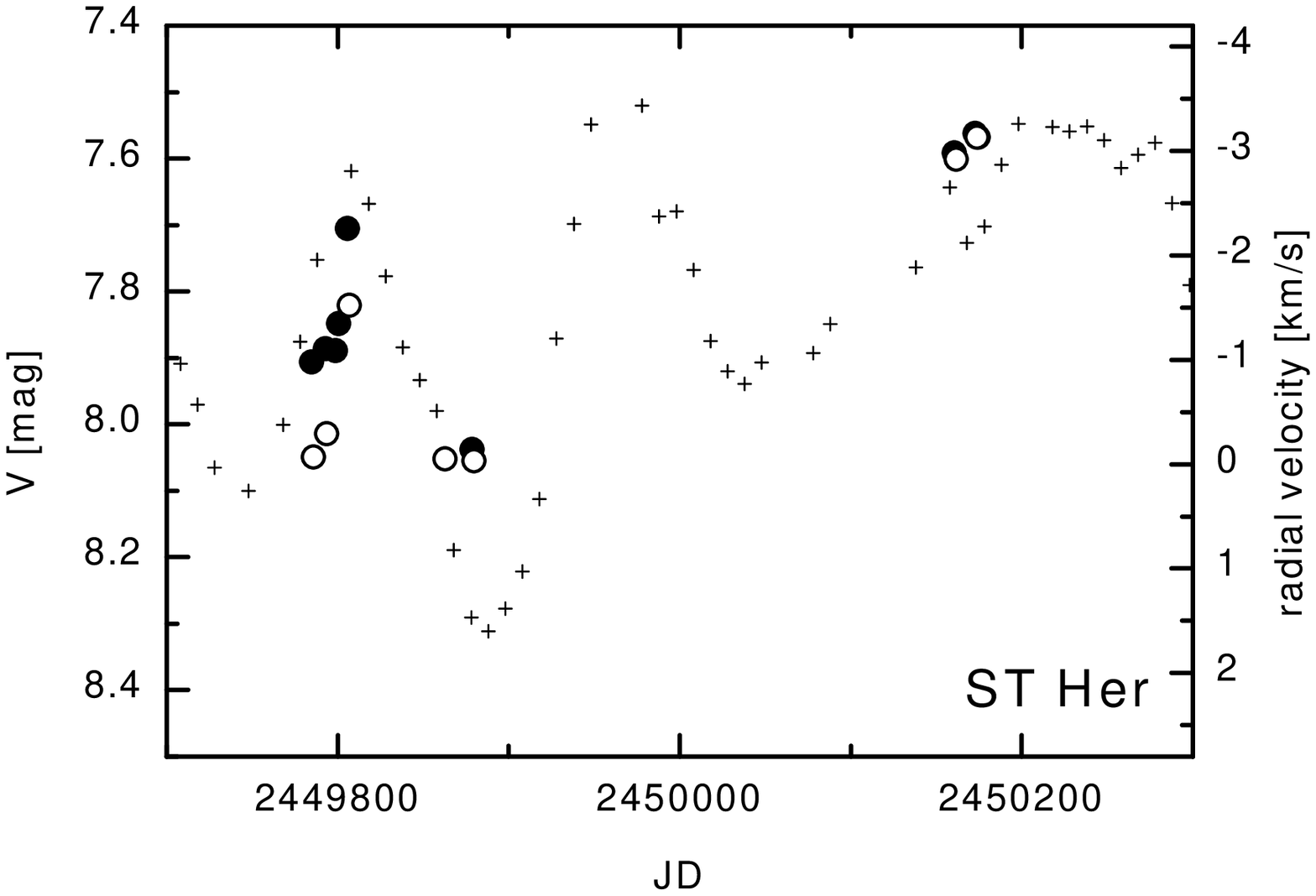}}
\caption[]{Light and velocity variations of ST Her. Stellar velocity
($-$22\,km/s from thermal microwave CO, see Lebzelter \cite{Lebzelter99})
has been subtracted from the velocity data. Same symbols as in Fig.~
\ref{fig6}. \label{fig8}} \end{figure}

\begin{figure} \resizebox{\hsize}!{\includegraphics{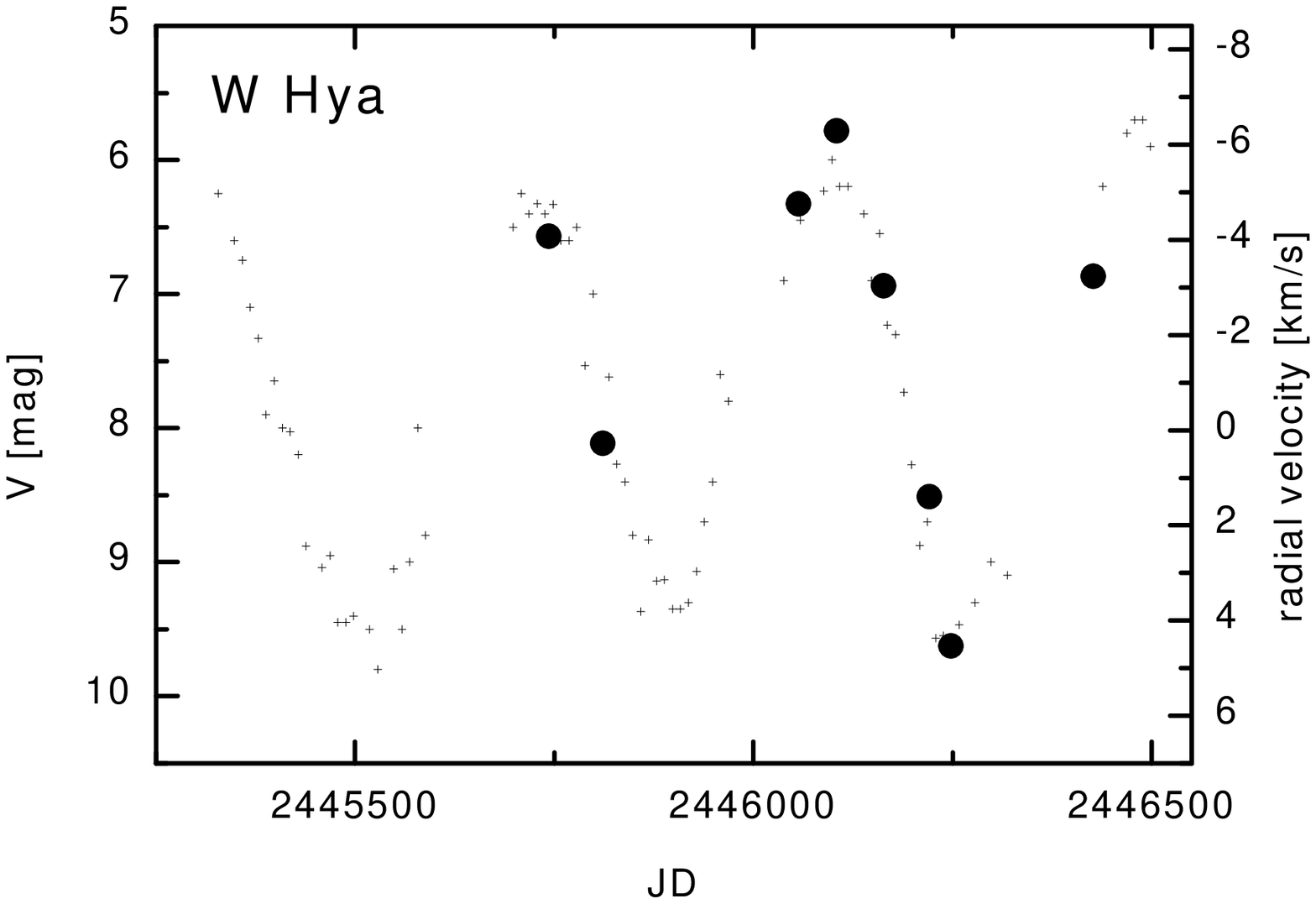}}
\caption[]{Light and velocity variations of W Hya. Stellar velocity 
(38\,km/s from thermal microwave CO, see Hinkle et al.~\cite{HLS97}) has
been subtracted from the velocity data. Same symbols as in Fig.~
\ref{fig1}. \label{fig9}} \end{figure}

In Figs.~\ref{fig1} to \ref{fig9} we plotted the
velocity variations on top of the light variations. For some stars
single velocity data points from a few years earlier exist. These were
not included in the plots to allow for a larger scale on the time axis.
A mean velocity of the star -- when available -- has been 
subtracted from
the velocity values. The velocity used for subtraction is given in the
figure caption. 
Lebzelter (\cite{Lebzelter99}) found that thermal CO velocities and other values
from the literature all appeared to share the same property, i.e.~a systematic
offset in the sense that the IR CO lines are blueshifted for all or a large part of
the lightcycle. We will come back to this point in the discussion of the results.
Note that some of these stellar velocities have
been derived from blue/optical spectra and may therefore not
be representative of the center-of-mass velocity. For miras it has
been shown that the blue/optical velocities are shifted several
km s$^{-1}$ positive of the center-of-mass velocity (Reid \cite{Reid76}, Hinkle \&
Barnes \cite{HB79}). Wallerstein \& Dominy (\cite{WD88}) found for 5 semiregular
variables without H$_\mathrm{2}$O masers that on the average there is
no velocity shift between optical and
thermal (=systemic) velocity. However, for 4 SRVs in their sample
showing H$_\mathrm{2}$O maser emission a difference
v$_\mathrm{thermal}$ $-$ v$_\mathrm{optical}$ of $-$1.3 km/s was found. 
It is therefore not clear whether the optical
velocity is representative for the center-of-mass velocity or not.

However, as thermal CO velocities are missing for many objects, the velocities used here
are the only external source of velocity information available. 
Keeping the above mentioned uncertainty in
mind we will use this velocity value as systemic velocity. We note that an investigation of
a systematic velocity shift between blue and thermal CO velocities is still missing for
semiregular variables. To avoid confusion we will add a remark on the source of the velocity
in the figure caption (thermal CO or blue/optical). For the reference of each velocity we refer to
the paper listed in Table\,\ref{tab1}.

The velocities have been plotted in reverse order
(largest velocity at the bottom) for the sake of an easier comparison
of velocity and light variations. Throughout this paper positive velocity 
relative to the
systemic velocity means that the material is moving toward the star and hence away from
the observer.
Furthermore the scale of the velocity
axis has been adjusted to have an overlap between velocity and light
curve.

It is obvious that the light and velocity variations are well
correlated in all objects of our sample.  Some -- but not all -- changes in
the light amplitude are well represented in the behavior of the
velocity curve as well. All stars reach their smallest difference to the systemic velocity 
at or
close to light minima, while the minimum velocities are correlated with
light maxima.

V450 Aql exhibits velocity changes with an amplitude of about
1.6\,km/s. This is the smallest variation in our sample and one of the
smallest among the stars for which Hinkle and collaborators monitored
CO velocities.  The accuracy of the velocity measurements is high
enough to ensure that the star is truly variable in velocity.
However, due to the small amplitude the uncertainties in the light curve
and the sampling of our velocity measurements
makes comparison of light and velocity curve rather difficult
(Fig.~\ref{fig1}).  Although it seems like V450 Aql behaves the same
way as all the other stars, even the opposite behavior (i.e.~maximum
velocity at maximum light) cannot be excluded.

Line doubling is present in the short period mira RT Cyg
(Fig.~\ref{fig5}) significantly before the visual light maximum. If we
compare the behavior of this star with the velocity curve of $\chi$
Cyg, we have to keep in mind that RT Cyg has a very symmetric light
curve compared to its long period counterpart. The slight phase shift
between the minimum velocity and the light maximum found in $\chi$ Cyg
is not visible in RT Cyg. However, it is possible that the phase
coverage is not adequate to detect such a small shift (which we would
expect to be only a few
days due to RT Cyg's short period).  Light curve data do not allow a
determination of the exact date of minimum around JD 2446200. Still the line
doubling observation with the maximum velocity observed seems to be
located clearly after the light minimum in agreement with the pattern seen in
$\chi$ Cyg. The single velocity measurement two cycles later fits
very well onto the light curve as well indicating that the correlation of
light and velocity curve is not limited to the well covered period.

The data for SV Cas (Fig.~\ref{fig2}) also merit further comments. The
behavior of this star was difficult to interpret without the light
curve. Plotting all data into one cyle made Hinkle et
al.~(\cite{HLS97}) suspect that the star might have a velocity curve
similar to the miras, i.e.~discontinuous. Our result shows that this
interpretation is not necessary, because the velocity variations follow
the semiregular light variations quite well. It was incomplete phase
coverage of the velocity observations that led to the
misinterpretation. The three maxima visible in Fig.~\ref{fig2} give a
mean period of only 237 days, somewhat less than the GCVS4 value used
by Hinkle et al.

The comparison of parallel velocity and light variations is of similar
importance for the understanding of the velocity variations in the case
of ST Her (Fig.~\ref{fig8}). Lebzelter (\cite{Lebzelter99})
assumed a phase shift between the velocity measurements around JD
2450150 and earlier data points to explain that the velocities did not
fit into a simple phase diagram. Fig.~\ref{fig8} shows that this
assumption was correct and that the star is not following its mean
period around that day. However, the velocities around that time fit
well onto the observed light curve.

For four additional stars (RV Boo, RR CrB, X Her, and TT Dra) 
velocities are available but good light
curves are not, making a detailed comparison, as for the stars in 
Table\,\ref{tab1}, impossible.  Using available light curves, we 
attempted to confirm that these stars behave as the stars
with good light curves.
All of them seem to exhibit the same kind
of correlation between light and velocity changes.

\section{Discussion} 

\subsection{The relation between light and velocity changes} 
From the figures it is obvious that the light and velocity changes in
semiregular variables are correlated. 
Most of the light curve irregularities,
especially variability in the length of a cycle, are physically linked
to a "semiregular" velocity curve. A relation
between the size of light and velocity change is indicated
by comparison of the velocity amplitude and light amplitude
listed in GCVS4 (Hinkle et al.~\cite{HLS97}). By using the light 
curves there is the possibility for a more direct comparison of these two
quantities.

Might it be possible to 
quantify the qualitative impression of the relation between 
light and velocity changes?  
Such an attempt is of course restricted by the limited accuracy both in
the light curve and in the velocity data, and the small time span
covered by the observations. Therefore we consider the following
result as preliminary.  The comparison of light and velocity amplitudes
by Hinkle et al.~(\cite{HLS97}) gives a mean ratio between light and velocity change
of approximately 4\,km/s/mag for the semiregular variables. However,
our comparison of simultaneously obtained velocity and light
measurements shows that the ratio ranges from 3 to 7\,km/s/mag.
(For V450 Aql and g Her we did not derive this quantity as velocity and
light changes do not correlate as well as the other
objects in our sample.) Most of the scatter will originate from the
lower time resolution of the velocity data leaving considerable
uncertainty in shifting and scaling the velocity curve onto the light
curve. While the order of the mean value of both investigations is the
same, the large scatter we found in the more detailed comparison
indicates a more complex relation between velocity and light change as
the latter will be influenced by other parameters, for example temperature
variations.  Data on bolometric light variations and
simultaneous information on the temperature change would be necessary
to allow a better quantification of such a relation.  Furthermore, due
to the obvious changes from cycle-to-cycle the definition of a mean
amplitude and an {\it amplitude range} will be the more appropriate
parameter.

Cummings et al.~(\cite{CHKG98}) presented high-precision relative
radial velocities measured in the optical wavelength region for 31 red
giants. They also made broad-band photometric observations to
differentiate between possible causes of velocity variations. The
observed light and radial velocity variations were either in inverted
phase, or shifted by a significant percentage (up to 30-40\%) of the
cycle-length.  What they found is in good agreement with our results.

\subsection{Indications for the cause of the variability} 
When identifying the mechanism responsible for the observed variations
two features visible in the data must be taken into account. First, all
SRVs show their maximum velocity at light minimum and their minimum
velocity at light maximum.  Second, most of the SRVs investigated here
show the puzzling feature that the measured CO velocities are all more negative then
the stellar velocity (Lebzelter \cite{Lebzelter99}).

\subsubsection{Velocity shift}
The observed velocity shift between the high excitation CO lines and the systemic
velocity has been mentioned already in Lebzelter (\cite{Lebzelter99}). For some stars
the velocity reference point has been obtained from blue-optical spectra. While it
is not clear whether this spectral range is useable to determine the systemic
velocity or not, we note that the same effect is observed for stars, for which the
center-of-mass velocity has been derived from microwave thermal CO observations.

Several mechanisms are thinkable for producing such a velocity shift. Some of them
have already been discussed by Lebzelter (\cite{Lebzelter99}). Pulsation may well
produce considerable velocity shifts and asymmetries between maximum outflow and infall 
velocity. It has been shown from model calculations of AGB stars
that traveling waves introduced by the pulsation can
persist longer than one light cycle in the stellar atmosphere 
(e.g.~H\"ofner \& Dorfi \cite{HD97}). These effects are correlated with the
strength of the pulsation. As the objects discussed here are mainly low amplitude
objects it is not clear yet, if a velocity shift of correct size can be produced
by pulsation alone. Models of late type pulsating stars calculated by Bowen 
(\cite{Bowen88}) indicate the
existence of a warm "chromospheric" layer of outflowing gas 
(called "calorisphere" by Willson \& Bowen \cite{WB86}).
If the CO lines we use in this investigation are formed in such a layer
then its outward motion may well explain the observed velocity shift. However,
the high excitation CO lines are quite weak, so it is more likely that they are
formed deep inside the stellar atmosphere. Furthermore, such a layer is not
found in the models by H\"ofner \& Dorfi (\cite{HD97}), therefore its existence is
still a matter of debate.
A further possible mechanism responsible for the velocity shift are large convective cells
on the stellar surface as suggested by Lebzelter (\cite{Lebzelter99}). However, detailed
modeling of variability due to stellar convection is not yet available for AGB stars.

In the 
following we will discuss the correlation between light and velocity changes in
the light of the possible reasons accounting for the velocity shift.

\subsubsection{Pulsation} 
In this section we discuss the case when the
pulsation is considered to be the dominant factor in producing stellar
light and velocity changes. In the case of variations driven only by
pulsation an assumption must be made to deal with the systematic difference
between the observations and the center-of-mass velocity. Possible
reasons have been mentioned above and in Lebzelter (\cite{Lebzelter99}).
However, the further discussion does not depend
on what reason is chosen. 

Superficially both the light curves and radial velocity curves are in 
agreement with
radial pulsation being the main reason for the light variations.
The observed properties, e.g. radial velocity minima occurring very
close to maximum light, are typical in radially pulsating stars
(RR~Lyrae-, high-amplitude $\delta$~Scuti- and Cepheid variables).  
In
radially pulsating stars the light- and radial velocity curves are in some cases
perfectly mirror images (see e.g.~Bersier et al.~\cite{BBB94a},
\cite{BBMD94b}). Our velocity curves are plotted in an
inverted form, therefore, they show exactly the same behavior as other
radial pulsators.  This means that the star reaches its maximum light
during the expansion.

However, such a 'mirror behavior' suggests that the light variation is
caused by radius variation only. From miras we know that 
the visual light change in late-type stars is dominated by the
temperature change via the variations in the strengths of the TiO
bands. The effect of the TiO bands is exactly the opposite way, leading
to a minimum flux (in the visual) when the star is largest, i.e.~has
its minimum temperature. In that case a phase shift between the CO line
forming layers and the layers producing the visual continuum is
necessary to explain the observations reported in this paper.  This is
in agreement with results from present AGB star models proposing
extended atmospheres where different spectral features can originate at
completely different radii from the stellar center.  Unfortunately, no
simultaneous temperature determinations exist for the SRVs of our sample. 

Accepting this pulsational approach, we estimated the order of
magnitude of the relative radius variation caused by the radial
motion.  A well observed example is W~Cygni.  W~Cygni's light 
curve is
dominated by two periods ($P_0=240$ days, $P_1=130$ days, Kiss et
al.~\cite{KSCM99a}).  These periods, their ratio and theoretical
calculations by Ostlie \& Cox (1986) imply the following stellar
parameters: $M=1.4 M_\odot$, $L=7000 L_\odot$, $T=3000 K$. The adopted
model gives $P_0=238$ days and $P_1=129$ days for the fundamental and
first overtone mode, respectively.  Recent models including the
coupling with convection (Xiong et al.~\cite{XDC98}) give $P_1=230$
days and $P_3=130$ days (first and third overtones) for $M=1.0
M_\odot$, $L=5000 L_\odot$ and $T=2700 K$. In either case the periods
fit for a stellar
radius of $\approx 300 R_\odot$. Considering the data subset around JD
2446000, the radial velocity of W~Cygni changed approximately 2
km~s$^{-1}$ in 70 days which gives an order-of-magnitude estimate for
the corresponding radial displacement about $\sim 17 R_\odot$. That is
5 percent relative radius change being a reasonable value for a radial
pulsator. This is, of course, a very oversimplified estimation.
We stress that several different layers with different
velocities might contribute to the final CO line profiles and hence 
to the measured velocities. Therefore radius variations calculated from
these velocity variations have to be taken with care (see
e.g.~Hinkle et al.~\cite{HHR82} for a discussion).  However, the most important
consequence of our simple estimate is that radial pulsation is a
realistic explanation for a set of semiregular variables.

\subsubsection{Convective cells} As a second approach we assume that
the outer atmospheric layers do not pulsate at all. The velocity
variations might then originate from convective motion. According to
Schwarzschild (\cite{Schwarzschild75}) and other authors only a small
number of convective cells cover the surface of an evolved red giant.
Lebzelter (\cite{Lebzelter99}) suggested that they might be responsible
for a systematic blueshift of the measured velocities. This is based on
the assumption that we mainly see the outflowing matter which is hotter
and therefore brighter than the matter flowing back onto the star. This
outflow has to be by no means constant from the (distant) observers
view.  While the individual motion within each granule on a star like
the sun will average out over the whole stellar disk, the situation is
different for the red giants due to the small number of cells on the
surface. Beside velocity variations within each individual cell the
movement of cells on the stellar surface as well as their evolution
will lead to significant variations both in light and velocity
integrated over the stellar disk.  Velocity variations caused by
convective motion are also in agreement with the observed coincidence
of maximum outflow velocity and light maximum. If the number of
convective cells on the surface increases, the area of matter falling
back onto the star, which forms the borders of these cells, increases.
Therefore the fraction of the surface covered by infalling matter
changes the brightness and the convective velocity shift in the same
way.  For a velocity variation within an individual convective cell the
same effect is achieved as a higher velocity will carry hot matter into
higher regions of the atmosphere leading to a brightening of the whole
star. To quantify this, detailed simulations are needed.
The time scale of the observed variations for the short period objects
is in agreement with the convective time scale of about 40 days as noted by Antia et
al.~(\cite{ACN84}). 

\section{Variability of semiregular variables}
The current data are not sufficient to determine the cause of
irregular variations in late type stars. These irregular variations
have been found in semiregular, irregular and mira-type variables.
The irregularities found in mira velocity curves are of similar size to the
velocity variations in the short period SRVs (see the data for
$\chi$\,Cyg in Hinkle et al.~\cite{HHR82}) and might have a common
cause.

We showed that both pulsation and convective cells can in principle
produce velocity and light changes. The standard pulsation interpretation
is in nice agreement with the observed correlation of velocity and light
changes. On the other hand it is very difficult to estimate the size
of the light and velocity variations due to convective cells without detailed
modelling. The interesting feature of convective cells is the irregular
variability and the systematic velocity shift they produce. 

There are several features found in semiregular variables that cannot be
explained without the assumption of stellar pulsation.
A number of SRVs like W Hya have periods a few times the timescale
estimated for convective cells. Furthermore, their rather
large velocity amplitude and the crossing of the center-of-mass
velocity by the velocity curve make an explanation with convective cells alone
for the long period objects rather unlikely.
It has also to be kept in mind that several semiregular variables at
least sometimes show hydrogen emission lines of the Balmer series in their
spectra. In the miras, and presumably in the SRVs, this emission 
component originates from shock waves in the stellar
atmospheres.  These shock waves are the
result of global stellar pulsation, giving a strong argument for a
pulsational origin for the light curve and velocity variations.

Models predict the existence of large convective cells at the surface of 
red giants. It is therefore very likely that the observed variability
in light and velocity includes a component provided by these surface
structures. This component will be found among small amplitude irregulars as well as among
miras and may account for the fact that many 'nonvariable' red giants are
actually variables with small amplitudes (see e.g.~Jorissen et al.~\cite{JMSM97}). 
If one wants to explain this
small amplitude variability with the help of large convective cells
it is critical to know the center-of-mass velocity very accurately.
Variability due to convective cells only should lead to a measured velocity smaller
than the center-of-mass velocity (i.e.~always outflow of matter) at all phases. 

It is well known that miras do not always reach the same brightness in each light
maximum. The differences (see e.g.~Fig.~\ref{fig5}) can be of the order of one magnitude.
Such 'irregularities' may be produced by processes on longer time scales than the
main stellar pulsation period (e.g.~H\"ofner \& Dorfi \cite{HD97}).
This illustrates that pulsation itself, even if only one mode is excited, can introduce
irregularities into the light change of long period variables.

At the moment it is not possible to estimate whether convective motion can
introduce the observed irregularity into the light and velocity changes 
alone or not. The variability of
the objects with the smallest amplitudes might be explained by
convective motion only. For those cases it does not seem necessary to
explain the small scale variability with the help of high degree
overtone modes as suggested by Percy \& Parkes (\cite{PP98}). However,
with the actual data available we cannot exclude regular pulsation or chaotic
behavior in these objects, but we want to support the possibility of
surface structures caused by convective motion as an alternative
explanation.

To summarize, we think it is a reasonable approach to see 
the observed variability in semiregular and irregular
variables as a composite of pulsation and an (additional?) irregularity
introduced by, e.g., large convective cells. The importance of each of
these two components will probably be very different for different objects.

A significant challenge for the late type variables is to define the
regions of instability on the H-R diagram.
If we want to have a chance to define the limiting parameters for the
onset of pulsation and to estimate the influence of convective motion on
the variability, a significantly larger sample of light and velocity
curves as well as other stellar characteristics (both fundamental as
well as atmospheric, for instance the occurrence of
emission lines) are needed. In this context the light curves provided
by amateur astronomers are of great value but they are not
accurate enough to avoid ambiguities in small light curve irregularities. 
Therefore only photoelectric data as provided e.g.~by
automatic telescopes can reveal these details in the light curves. Both
sources of data are needed to search for possible further components
contributing to the light change of semiregular and irregular
variables.

\begin{acknowledgements} 
This work was supported by Austrian Science
Fund Project S7308 and Hungarian OTKA Grants T022259 and F022249.  This
research made use of the SIMBAD database operated by CDS in Strasbourg,
France.  In this research, we have used, and acknowledge with thanks,
data from the AAVSO International Database, based on observations
submitted to the AAVSO by variable star observers worldwide.  We also
thank similar computer services maintained by the extremely
enthusiastic staff of the AFOEV and VSOLJ.  \end{acknowledgements}

\end{document}